\documentclass[twoside,twocolumn,9pt]{article}
\usepackage{extsizes}
\usepackage[super,sort&compress,comma]{natbib} 
\usepackage[version=3]{mhchem}
\usepackage[left=1.5cm, right=1.5cm, top=1.785cm, bottom=2.0cm]{geometry}
\usepackage{balance}
\usepackage{mathptmx}
\usepackage{sectsty}
\usepackage{graphicx} 
\usepackage{lastpage}
\usepackage[format=plain,justification=justified,singlelinecheck=false,font={stretch=1.125,small,sf},labelfont=bf,labelsep=space]{caption}
\usepackage{float}
\usepackage{fancyhdr}
\usepackage{fnpos}
\usepackage[english]{babel}
\addto{\captionsenglish}{%
  
}
\usepackage{array}
\usepackage{droidsans}
\usepackage{charter}
\usepackage[T1]{fontenc}
\usepackage[usenames,dvipsnames]{xcolor}
\usepackage{setspace}
\usepackage[compact]{titlesec}
\usepackage{hyperref}

\definecolor{cream}{RGB}{222,217,201}

\begin{document}

\pagestyle{fancy}
\thispagestyle{plain}
\fancypagestyle{plain}{
\renewcommand{\headrulewidth}{0pt}
}

\makeFNbottom
\makeatletter
\renewcommand\LARGE{\@setfontsize\LARGE{15pt}{17}}
\renewcommand\Large{\@setfontsize\Large{12pt}{14}}
\renewcommand\large{\@setfontsize\large{10pt}{12}}
\renewcommand\footnotesize{\@setfontsize\footnotesize{7pt}{10}}
\makeatother

\renewcommand{\thefootnote}{\fnsymbol{footnote}}
\renewcommand\footnoterule{\vspace*{1pt}%
\color{cream}\hrule width 3.5in height 0.4pt \color{black}\vspace*{5pt}} 
\setcounter{secnumdepth}{5}

\makeatletter 
\renewcommand\@biblabel[1]{#1}            
\renewcommand\@makefntext[1]%
{\noindent\makebox[0pt][r]{\@thefnmark\,}#1}
\makeatother 
\renewcommand{\figurename}{\small{Fig.}~}
\sectionfont{\sffamily\Large}
\subsectionfont{\normalsize}
\subsubsectionfont{\bf}
\setstretch{1.125} 
\setlength{\skip\footins}{0.8cm}
\setlength{\footnotesep}{0.25cm}
\setlength{\jot}{10pt}
\titlespacing*{\section}{0pt}{4pt}{4pt}
\titlespacing*{\subsection}{0pt}{15pt}{1pt}

\fancyfoot{}
\fancyfoot[LO,RE]{\vspace{-7.1pt}\includegraphics[height=9pt]{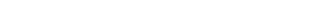}}
\fancyfoot[CO]{\vspace{-7.1pt}\hspace{13.2cm}\includegraphics{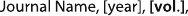}}
\fancyfoot[CE]{\vspace{-7.2pt}\hspace{-14.2cm}\includegraphics{RF}}
\fancyfoot[RO]{\footnotesize{\sffamily{1--\pageref{LastPage} ~\textbar  \hspace{2pt}\thepage}}}
\fancyfoot[LE]{\footnotesize{\sffamily{\thepage~\textbar\hspace{3.45cm} 1--\pageref{LastPage}}}}
\fancyhead{}
\renewcommand{\headrulewidth}{0pt} 
\renewcommand{\footrulewidth}{0pt}
\setlength{\arrayrulewidth}{1pt}
\setlength{\columnsep}{6.5mm}
\setlength\bibsep{1pt}

\makeatletter 
\newlength{\figrulesep} 
\setlength{\figrulesep}{0.5\textfloatsep} 

\newcommand{\topfigrule}{\vspace*{-1pt}%
\noindent{\color{cream}\rule[-\figrulesep]{\columnwidth}{1.5pt}} }

\newcommand{\botfigrule}{\vspace*{-2pt}%
\noindent{\color{cream}\rule[\figrulesep]{\columnwidth}{1.5pt}} }

\newcommand{\dblfigrule}{\vspace*{-1pt}%
\noindent{\color{cream}\rule[-\figrulesep]{\textwidth}{1.5pt}} }

\makeatother

\twocolumn[
  \begin{@twocolumnfalse}
{\includegraphics[height=30pt]{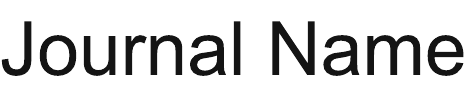}\hfill\raisebox{0pt}[0pt][0pt]{\includegraphics[height=55pt]{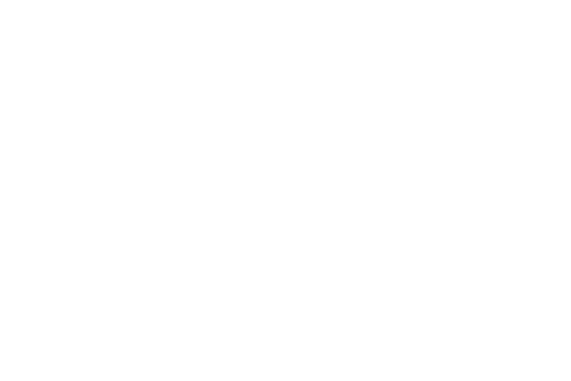}}\\[1ex]
\includegraphics[width=18.5cm]{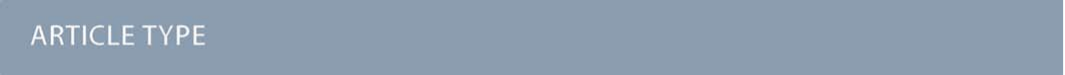}}\par
\vspace{1em}
\sffamily
\begin{tabular}{m{4.5cm} p{13.5cm} }

  \includegraphics{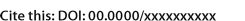} & \noindent\LARGE{\textbf{The diverse chemistry of protoplanetary disks as revealed by JWST}}\\
\vspace{0.3cm} & \vspace{0.3cm} \\

 & \noindent\large{Ewine F. van Dishoeck,$^{\ast}$\textit{$^{a,b}$} S. Grant\textit{$^{b}$}, B. Tabone\textit{$^{c}$}, M. van Gelder\textit{$^{a}$}, L. Francis\textit{$^{a}$}, L.\ Tychoniec\textit{$^{d}$}, G. Bettoni\textit{$^{b}$}, A.M. Arabhavi\textit{$^{e}$}, D. Gasman\textit{$^{f}$}, P. Nazari\textit{$^{a}$}, M. Vlasblom\textit{$^{a}$}, P. Kavanagh\textit{$^{g}$}, V. Christiaens\textit{$^{h}$}, P. Klaassen\textit{$^{i}$}, H. Beuther\textit{$^{j}$}, Th. Henning\textit{$^{j}$}, and I. Kamp\textit{$^{e}$} } \\

  \includegraphics{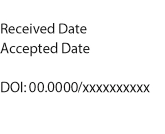} & \noindent\normalsize{Early results from the JWST-MIRI guaranteed time programs on protostars (JOYS) and disks (MINDS) are presented. Thanks to the increased sensitivity, spectral and spatial resolution of the MIRI spectrometer, the chemical inventory of the planet-forming zones in disks can be investigated with unprecedented detail across stellar mass range and age. Here data are presented for five disks, four around low-mass stars and one around a very young high-mass star. The mid-infrared spectra show some similarities but also significant diversity: some sources are rich in CO$_2$, others in  H$_2$O or C$_2$H$_2$. In one disk around a very low-mass star, booming C$_2$H$_2$ emission provides evidence for a ``soot'' line at which carbon grains are eroded and sublimated, leading to a rich hydrocarbon chemistry in which even di-acetylene (C$_4$H$_2$) and benzene (C$_6$H$_6$) are detected\cite{Tabone23}. Together the data point to an active inner disk gas-phase chemistry that is closely linked to the physical structure (temperature, snowlines, presence of cavities and dust traps) of the entire disk and which may result in varying CO$_2$/H$_2$O abundances and high C/O ratios $>$1 in some cases. Ultimately, this diversity in disk chemistry will also be reflected in the diversity of the chemical composition of exoplanets.}
  \\

\end{tabular}

 \end{@twocolumnfalse} \vspace{0.6cm}

  ]

\renewcommand*\rmdefault{bch}\normalfont\upshape
\rmfamily
\section*{}
\vspace{-1cm}


\footnotetext{\textit{$^{a}$Leiden Observatory, Leiden University, P.O. Box 9513, 2300 RA Leiden, the Netherlands. E-mail: ewine@strw.leidenuniv.nl}}
\footnotetext{\textit{$^{b}$Max-Planck Institut f\"{u}r Extraterrestrische Physik (MPE), Giessenbachstr.\ 1, 85748, Garching, Germany. }}
\footnotetext{\textit{$^{c}$Universit\'e Paris-Saclay, CNRS, Institut d’Astrophysique Spatiale, 91405, Orsay, France. }}
\footnotetext{\textit{$^{d}$European Southern Observatory, Karl-Schwarzschild-Strasse 2, 85748 Garching bei M\"unchen, Germany. }}
\footnotetext{\textit{$^{e}$Kapteyn Astronomical Institute, Rijksuniversiteit Groningen, P.O. Box 800, 9700 AV Groningen, The Netherlands. }}
\footnotetext{\textit{$^{f}$Institute of Astronomy, KU Leuven, Celestijnenlaan 200D, 3001 Leuven, Belgium. }}
\footnotetext{\textit{$^{g}$Dublin Institute for Advanced Studies, Astronomy \& Astrophysics Section, 31 Fitzwilliam Place, Dublin 2, Ireland. }}
\footnotetext{\textit{$^{h}$STAR Institute, Universit\'e de Li\`ege, All\'ee du Six Ao\^ut 19c, 4000 Li\`ege, Belgium. }}
\footnotetext{\textit{$^{i}$UK Astronomy Technology Centre, Royal Observatory Edinburgh, Blackford Hill, Edinburgh EH9 3HJ, UK. }}
\footnotetext{\textit{$^{j}$Max-Planck-Institut f\"{u}r Astronomie (MPIA), K\"{o}nigstuhl 17, 69117 Heidelberg, Germany. }}





\section{Introduction}

Planets are built from gas and solids in the rotating disks around
young stars. Their composition thus depends on the gas-grain chemistry
that takes place in disks, which is known to change with position due
to strong gradients in temperature, density and UV irradiation (e.g.,
refs.\cite{Oberg21,vanDishoeck21exo,Miotello22}). This chemistry may
also change with evolutionary state: it is now well established that
disks form early, already in the embedded state of star formation when
the disk is still warmer \cite{vantHoff20} and surrounded by a larger
scale envelope that can feed the disk with fresh material. Planetary
cores are likely formed at these early stages as well
\cite{Tychoniec20}. The study of the chemistry of planet formation
must therefore encompass the entire sequence of star formation, from
cold collapsing clouds to warm circumstellar disks.

The {\it James Webb Space Telescope} (JWST) offers new opportunities
to study the chemistry of the warm inner regions of disks, typically
within a few au$^*$ radius from a solar-type star. This inner disk is
the region in which most planets -- both gas giants and terrestrial
planets -- are thought to form \cite{Morbidelli12,Dawson18}. Compared
with the {\it Spitzer} Space Telescope, which revealed that inner
disks can have a very rich chemistry
\cite{Carr08,Salyk11,Pontoppidan14}, the JWST spectrometers have
higher spectral resolving power
($R=\lambda/\Delta \lambda \approx 2000-4000$ versus $60-600$) and
more than two orders of magnitude higher sensitivity, combined with an
order of magnitude higher spatial resolution.  JWST also complements
the Atacama Large Millimeter/submillimeter Array (ALMA) which resolves
millimeter-sized dust emission in disks down to a few au but which probes
chemistry primarily beyond $\sim$10 au (e.g.,
refs.\cite[]{Cleeves18IAU,Oberg21MAPS}).  \footnotetext{$^*$ 1
  astronomical unit (au) = distance Sun--Earth = 1.5$\times 10^{13}$
  cm}

We present here some first results of the chemistry in the inner
regions of disks based on JWST data from the Medium Resolution
Spectrometer (MRS) \cite{Wells15} that is part of the Mid-InfraRed
Instrument (MIRI) \cite{Rieke15,Wright23}. The data come from two
guaranteed time programs: the ``JWST Observations of Young
protoStars'' (JOYS) program (PIs: E.F.\ van Dishoeck and H.\ Beuther,
program id: 1290), which targets two dozen young disks and their
envelopes and outflows in the embedded stage of low- and high-mass
star formation \cite{vanDishoeck23}, and the MIRI INfrared Disk Survey
(MINDS) (PIs: Th.\ Henning and I.\ Kamp, program id: 1282), which
observes about 50 disks around pre-main sequence stars across a range
of stellar masses and ages \cite{Kamp23,Henning23}. Here high
  mass refers to stars of O and B spectral type (typical masses 8
  M$_{\odot}$ and higher) or protostars with total luminosities
  greater than a few $\times 10^3$ L$_\odot$. At the time of
submission of this paper (January 2023), only one JOYS source has been
observed ---the high-mass protostar IRAS 23385+6052
\cite{Cesaroni19}---, and data for a dozen MINDS sources have been
taken, although none of them are disks that were found to be
particularly line rich with {\it Spitzer} like AA Tau
\cite{Carr08}. Here we present results on small molecules, most
notably CO$_2$, H$_2$O and hydrocarbons, as observed toward IRAS
23385+6052, and from disks around the T Tauri stars GW Lup, V1094 Sco,
Sz 98 and the very low-mass star 2MASS-J16053215-1933159. Even within
this limited sample observed so far, the data show the diversity of
spectra and disk chemistry, not just among T Tauri disks but also
across a range of stellar masses.

\section{Observations and methods}

\subsection{JWST data reduction}

The MIRI-MRS consists of four Integral Field Units (IFUs) (also called
``channels'') covering different wavelength ranges within the
4.9--28.1 $\mu$m (355--2040 cm$^{-1}$) range.  The Field-of-View (FOV)
varies between the four channels from $3.2'' \times 3.7''$ covered
with 21 IFU slices of 0.18$''$ at the shortest wavelengths, to
$6.6''\times 7.7''$ observed with 12 slices of 0.66$''$ width at the
longest wavelengths.  Thus, the MIRI-MRS provides spectral images on
arcsec scales but in this paper only point-source spectra centered on
the disks are presented. Full wavelength coverage was obtained in
three grating moves
that are observed simultaneously for all four channels. The FASTR1
readout pattern was used with a 2-point dither pattern in JOYS and a
4-point dither pattern in MINDS.

IRAS 23385+6052 (luminosity $L$=$3\times 10^3$ L$_\odot$, distance
$d$=4.9 kpc$^\dag$ \cite{Molinari98}, age few$\times 10^4$ yr,
hereafter IRAS 23385) was observed as part of JOYS on August 22,
2022. This source actually consists of a small cluster of
  protostars, as revealed from previous millimeter
  \cite{Beuther18,Cesaroni19} and near-infrared
  images\cite{Faustini09}, with the most massive protostar estimated
  to be 9 M$_\odot$ from Keplerian rotation measurements of the
  surrounding gas \cite{Cesaroni19}. Note that this source is too far
  north to be observable by ALMA. The total exposure time in each MRS
grating setting was 200 sec.  \footnotetext{$^\ddag$ 1 parsec = 206265
  au = 3.26 lightyear = 3.086$\times 10^{18}$ cm}

The spectra toward GW Lup ($M_*$=0.46 M$_\odot$, $L$=0.33 L$_{\odot}$
\cite{Alcala17}, $d$=155 pc), V1094 Sco ($M_*\approx$0.9 M$_\odot$,
$L$=1.7 L$_{\odot}$ \cite{Alcala17}, $d$=153 pc), Sz 98
($M_*\approx$0.74 M$_\odot$, $L$=1.5 L$_{\odot}$ \cite{Alcala17},
$d$=155 pc), and 2MASS-J16053215-1933159 ($M_*$=0.14 M$_\odot$,
$L$=0.04 L$_{\odot}$ \cite{Luhman18}, $d$=152 pc, hereafter J160532)
were taken as part of the MINDS program on 2022 August 8 and August 1
for a total of $\sim$2 hr each including overheads (typically
800--1200 sec per grating setting). The distances to these
sources come from the Gaia DR3 catalogue \cite{GaiaDR3}, whereas their
ages are all typically a few Myr \cite{Alcala17,Miret22}.

The MIRI-MRS observations of all JOYS and MINDS sources were processed
through the three reduction stages \cite{Bushouse22} using Pipeline
version 1.8.4 of the JWST Science Calibration Pipeline43 and the CRDS
context {\tt jwst$_-$1017.pmap}. Prior to processing J160532, the
background was removed using pair-wise dither subtraction.  The
default class {\tt Spec2} was applied, skipping the residual fringe
correction.  Instead, for GW Lup, V1094 Sco, and Sz 98, the
alternative reference files as discussed in Gasman et
al.\cite{Gasman23calib} were applied, which correct the fringes
sufficiently in {\tt Spec2}. The data were then processed by the {\tt
  Spec3} class, which combines the calibrated data from the different
dither observations into a final level3 spectral cube. The {\tt
  outlier$_-$detection} and {\tt master$_-$background} methods were
skipped, since for both MINDS and JOYS this was done manually.
Finally, a manual extraction of the spectra for each sub-band was
performed, including aperture correction, with an aperture size of
2.5$\lambda / D$ for the low-mass sources.  The backgrounds of
GW Lup, V1094 Sco, and Sz 98 were estimated using an annulus around
the source, with an appropriate aperture correction accounting for the
PSF signal present in the annulus.

The high-mass protostar IRAS 23385 turns out to be a binary source
with a separation of $\sim$0.67$''$ ($\sim 3280$ au) at mid-infrared
wavelengths. This means that the binary is resolved in Channel 1 and
part of Channel 2, but that the system is unresolved at longer
wavelengths. The spectrum presented here was extracted using a
circular aperture of $2.4"$ diameter independent of wavelength
that includes both sources.  Background subtraction was done using the
emission measured at off-source positions within the IFUs.  This
turned out to be more reliable than subtraction using the
complementary dark field that was observed but that still contained
strong extended emission.  A residual fringe correction was applied at
the spectrum level. More details for IRAS 23385 can be found in
Beuther et al.\cite{Beuther23}, and for the MINDS sources in Grant et
al.\cite{Grant23} and Tabone et al.\cite{Tabone23}.

\subsection{Methods}

The mid-infrared wavelength range observed by MIRI-MRS covers the
vibration-rotation bands of many molecules, from
simple molecules like CO and CO$_2$ to more complex aliphatic and
aromatic hydrocarbon molecules. For H$_2$O and OH, also pure
rotational transitions between high-lying levels within the ground
vibrational state are observable beyond 10 $\mu$m.

The continuum-subtracted spectra were fitted with slab models of the
molecular emission that take optical depth effects and line overlap
into account, following the methods used for analyzing {\it Spitzer}
spectra \cite{Carr11,Salyk11}. The populations of the molecular levels
are assumed to be characterized by a single excitation temperature,
$T_{\rm ex}$, which is taken to be equal to the gas temperature $T$
assuming Local Thermodynamic Equilibrium (LTE).  Note that radiative
pumping of the lines may also be significant in the inner disk
\cite{Bruderer15}, in which case $T_{\rm ex}$ approaches the
temperature of the radiation field, $T_{\rm rad}$.

The line profile function is taken to be Gaussian with a full width at
half maximum of $\Delta V$=4.7 km s$^{-1}$ ($\sigma$=2 km s$^{-1}$),
as in previous {\it Spitzer} studies.  At high optical depth,
the line shape becomes a flat-topped pseudo-Gaussian due to saturation
at the line core.  These models have only three free parameters: the
line of sight column density $N$, the gas temperature $T$, and the
emitting area given by $\pi R^2$. It is stressed that $R$ does not
necessarily correspond to a disk radius, but that the emission could
come from a ring or any other region with an area equivalent to
$\pi R^2$.

\begin{figure}[tbh]
\centering
  \includegraphics[width=0.48\textwidth]{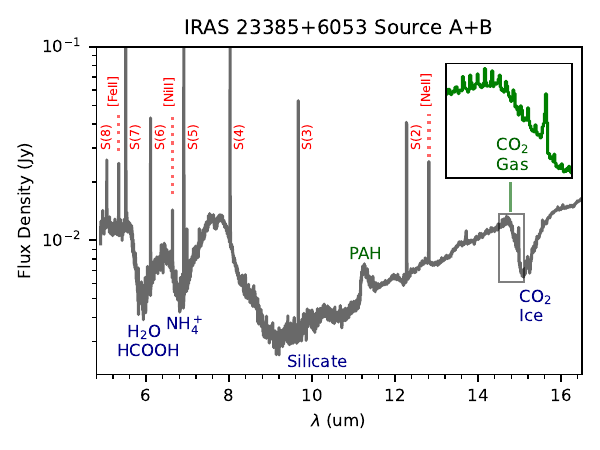}
  \caption{Part of the MIRI-MRS spectrum toward the high-mass
    protostar IRAS 23385+6053 integrated over both sources
  . Blue labels refer to ice and solid-state absorption features, red
  to spatially extended narrow emission lines, and green to 
    molecular emission with CO$_2$ centered on the sources and PAH
    emission extended.}
  \label{Fig1}
\end{figure}

\begin{figure}[tbh]
\centering
  \includegraphics[width=0.48\textwidth]{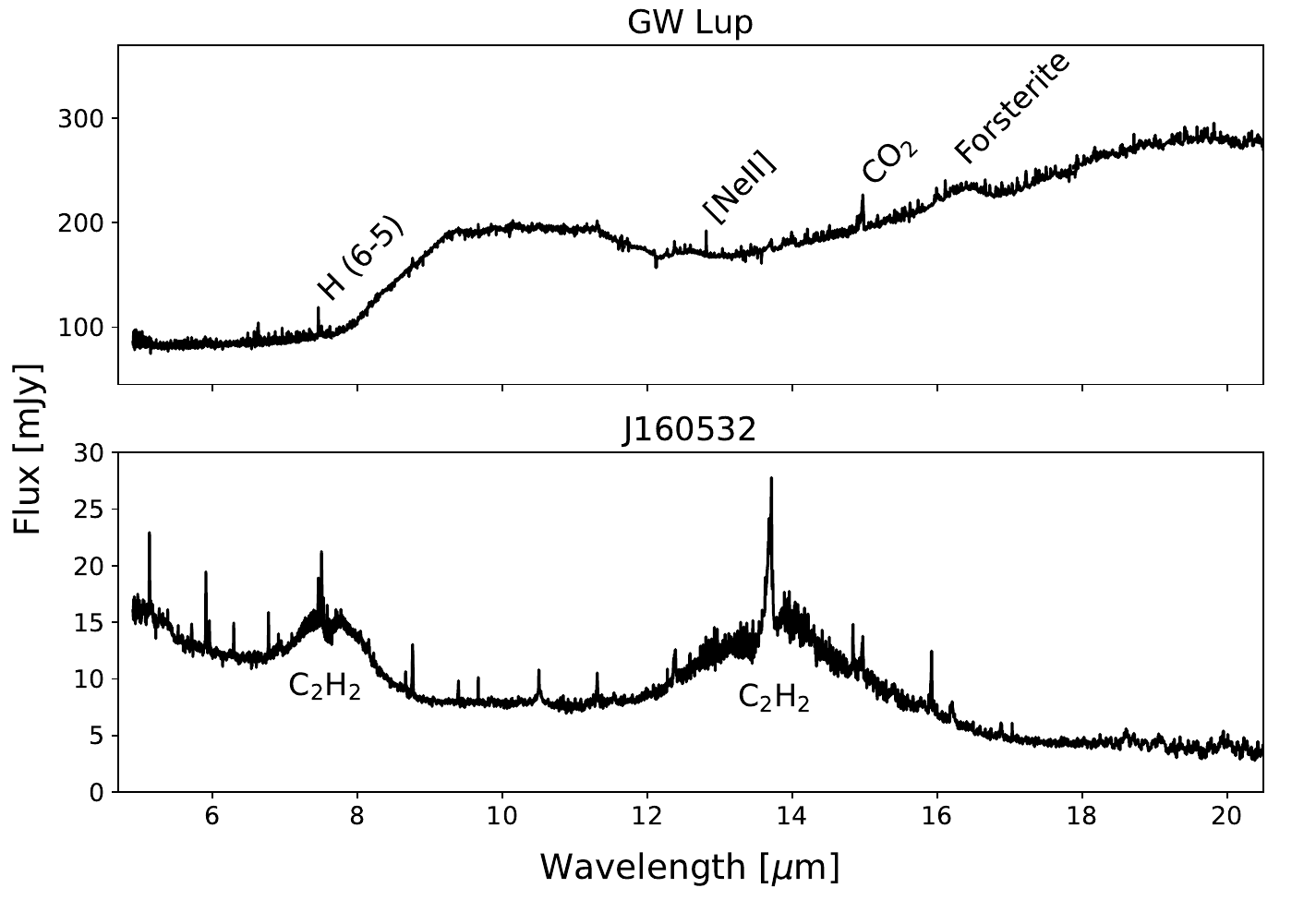}
  \caption{MIRI-MRS 5--20 $\mu$m spectrum of the disk around the T
    Tauri star GW Lup \cite{Grant23} compared with that of the
      very low-mass star J160532 presented by
      ref.\cite{Tabone23}. The GW Lup spectrum shows amorphous
    silicate emission at 10 and 18 $\mu$m with superposed atomic and
    molecular lines as well as some crystalline forsterite
    emission. The broadband J160532 spectrum shows no silicate
    emission but two prominent bumps that can be ascribed to very
    abundant C$_2$H$_2$ emission \cite{Tabone23}.}
  \label{Fig2}
\end{figure}

The molecular data, i.e., energy levels, statistical weights, Einstein
$A$ coefficients and partition functions, were taken from the HITRAN
2020 database \cite{Gordon22} and converted into LAMDA format
\cite{vanderTak20}. For C$_6$H$_6$, molecular parameters based on the
GEISA database were used \cite{GEISA21}.
Integrated line fluxes are computed, varying the emitting area and
taking the known distances to the sources into account.  The model
spectrum is then convolved to the instrumental spectral resolving power
\cite{Labiano21} and the convolved model spectrum is resampled to have
the same wavelength grid as the observed spectrum.  Model grids were
run for each molecule with $T$ from 100 to 1500 K, $N$ from 10$^{14}$
to 10$^{22}$ cm$^{-2}$, and emitting radius from 0.01 to 10 au for the
low-mass sources. For the high-mass source, these ranges are 50--550
K, 10$^{14}$--10$^{20}$ cm$^{-2}$ and 0.01--10000 au. The best-fit $N$
and $T$ are determined using a $\chi^2$ fit between the continuum
subtracted data and the convolved and resampled model spectrum, using
the best-fit emitting area for each $N$ and $T$. The procedure is
iterative, starting with fitting those molecules that have the most
lines in a certain wavelength range that are overlapping with those of others,
and ending with the weaker features. More details can be found in
refs.\cite{Tabone23,Grant23}.

\section{Results}

\subsection{Broad-band spectra}

Figure~\ref{Fig1} presents part of the observed MIRI-MRS spectrum
toward the high-mass binary protostar IRAS 23385
\cite{Beuther23}. Note that this source is more than a factor of
  1000 weaker than the high-mass protostars studied with ISO
  \cite{Gibb04}, only $\sim$10 mJy, demonstrating the JWST
  sensitivity. As is typical for a deeply embedded young stellar
object, the spectrum is rising with wavelength due to the thermal
emission from dust surronding the binary protostar which has a
temperature gradient from warm to cold \cite{Adams87}. Superposed on
this continuum are deep and broad absorption bands due to silicates
and a variety of ices arising in the cold outer envelope
\cite{Gibb00,Boogert15}. A number of strong, narrow emission lines due
to various atoms and H$_2$ are seen as well, associated with shocks
due to the outflows from the protostars. Finally, PAH emission from a
background cloud is seen at 11.3 and 8.6 $\mu$m \cite{Molinari08}; the
shorter wavelength PAH features are blocked by the high extinction and
deep ice features in the envelope. Both the absorptions and spatially
extended gas-phase lines will be described elsewhere
\cite{Beuther23,Gieser23}.

\begin{figure*}[h]
\centering
  \includegraphics[width=0.9\textwidth]{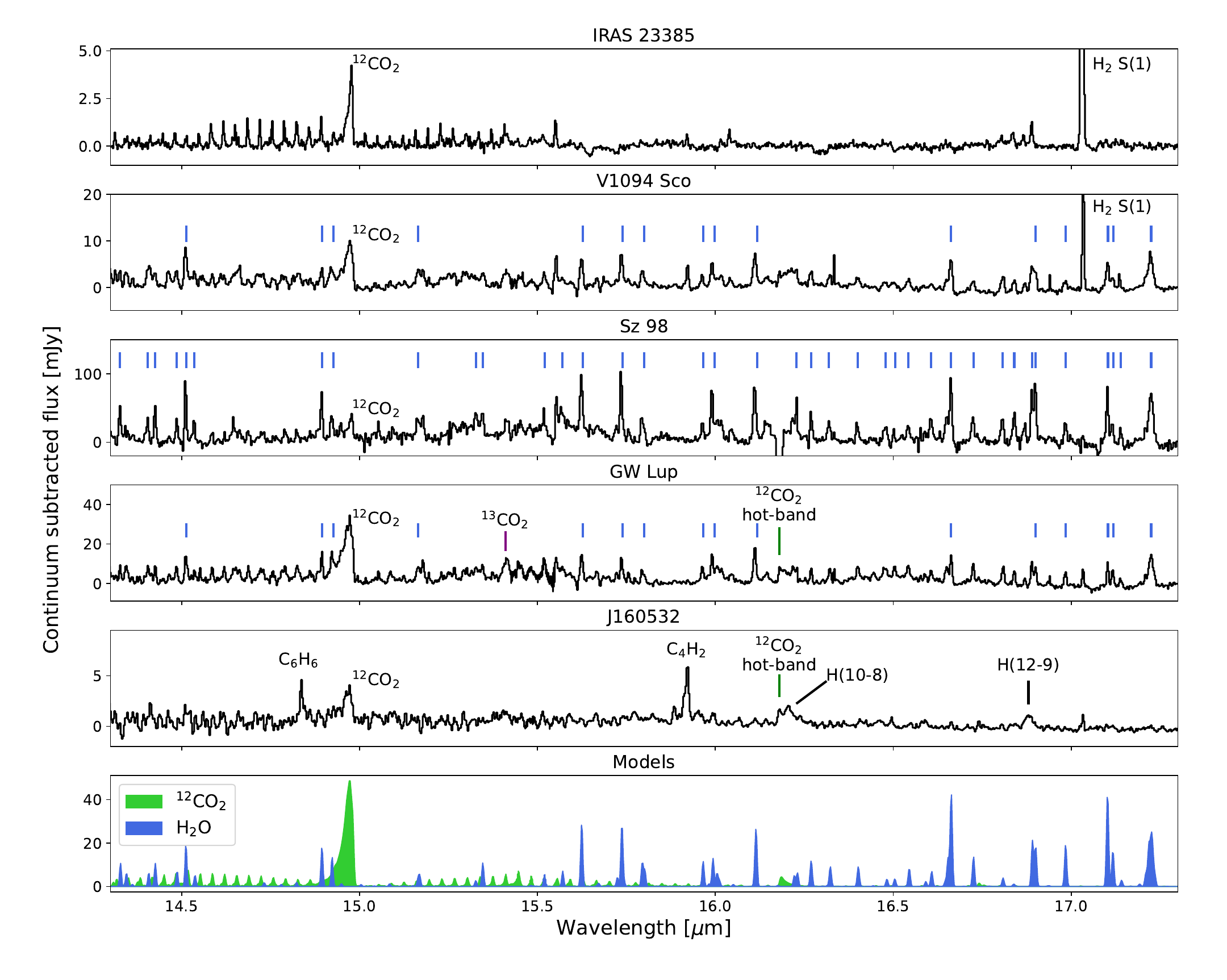}
  \caption{Continuum-subtracted MIRI-MRS 14.3--17.3 $\mu$m spectra of
    the young high-mass protostar IRAS 23385 \cite{Francis23}, and of
    the disks around the T Tauri stars V1094 Sco, Sz 98
    \cite{Gasman23} and GW Lup \cite{Grant23}, and the very low-mass
    star J160532\cite{Tabone23}. The blue tickmarks in the V1094 Sco,
    Sz 98 and GW Lup panels indicate the locations of water
    lines. Note also the detected $^{13}$CO$_2$ and hot bands of
    $^{12}$CO$_2$ toward GW Lup \cite{Grant23}. The bottom panel shows
    simulated LTE slab model spectra of CO$_2$ and H$_2$O at 600 K (arbitrary
    relative scaling).}
  \label{Fig3}
\end{figure*}

Figure~\ref{Fig2} compares the MIRI-MRS spectra of two low-mass
sources, GW Lup and J160532. The GW Lup spectrum is typical of that of
a disk around a young star whose protostellar envelope has already
been dissipated: its spectral energy distribution
drops at long wavelengths but it shows strong, broad
emission bands due to amorphous and crystalline silicates
\cite{Kessler-Silacci06,Furlan06,Henning10} in the MIRI range.

In contrast to most disk sources including V1094 Sco and Sz 98, the
J160532 spectrum is unusual: it shows no silicate emission feature but
has two broad bumps centered at 7.7 and 13.7 $\mu$m as was already
noted in low-resolution {\it Spitzer} data \cite{Dahm09,Pascucci13}.  Tabone
et al.\ \cite{Tabone23} show that these bumps are due to the
$\nu_4 + \nu_5$ and $\nu_5$ bands of C$_2$H$_2$ respectively which is
present at very high column densities ($\sim 10^{21}$ cm$^{-2}$) and
emitting at a temperature around 500 K within a small region of radius
0.033 au.  In this optically thick component, the $Q-$branch is
suppressed. The clearly visible $Q-$ branch as well as other
individual $R-$ and $P-$lines (see below) require an additional
optically thin C$_2$H$_2$ component with a 10$^4$ times lower column
density.

None of these disk spectra shows emission from polycyclic aromatic
hydrocarbons (PAHs). However, this lack of PAH features does not imply
absence of these molecules: as shown by Geers et al.\ \cite{Geers06},
the UV radiation from stars with $T_{\rm eff} < 4200$ K (spectral type
later than K6, assuming no significant excess UV radiation due to
accretion) is too weak to produce detectable PAH bands.

\begin{figure*}[tbh]
\centering
  \includegraphics[width=0.7\textwidth]{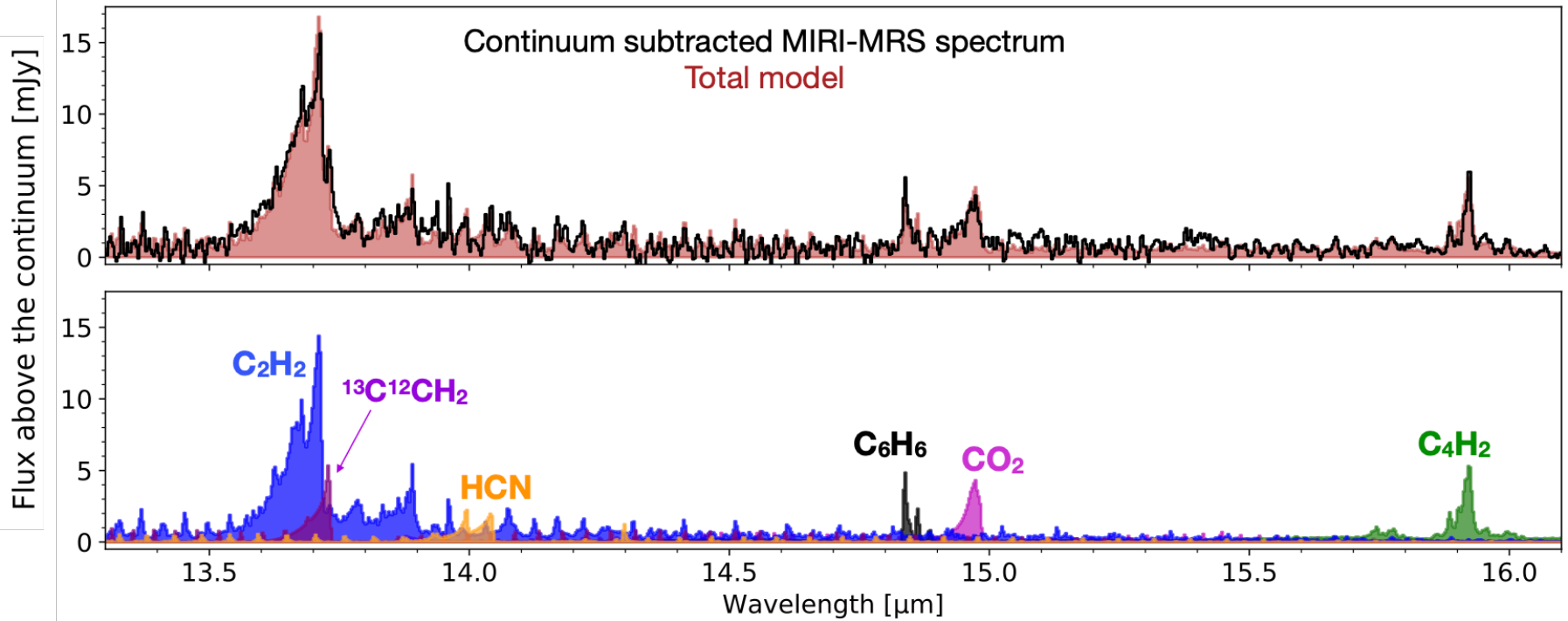}
  \caption{Continuum-subtracted MIRI-MRS 13.3--16.1 $\mu$m spectrum of the
    very low-mass star J160532, showing prominent features of
    C$_2$H$_2$, C$_4$H$_2$ and C$_6$H$_6$. The 13-C isotopolog of
    C$_2$H$_2$ is also detected, as are HCN and CO$_2$ (see
      ref.\cite{Tabone23} for details). The big bump at 13.7 $\mu$m
    (see Figure~\ref{Fig2}) due to very high column density C$_2$H$_2$
    component has been subtracted.}
  \label{Fig4}
\end{figure*}

\subsection{Identified molecules}

Of interest to this paper are the series of weak narrow lines with a
line/continuum ratio of less than 10 \% on top of the continuum (see
GW Lup spectrum in Figure~\ref{Fig2}). Figure~\ref{Fig3} presents the
continuum subtracted spectra in the 14.3--17.3 $\mu$m range for the
five sources discussed here, with various emission features
labelled. Fig.~\ref{Fig4} presents the spectrum of J160532 at
13.3--16.1 $\mu$m including the prominent C$_2$H$_2$ $Q-$branch at 13.7
$\mu$m with weak HCN at 14.0 $\mu$m. The shorter wavelength parts of
the spectra of the other sources showing C$_2$H$_2$ and HCN are
included in the paper by Kamp et al.\ (this volume).

The most prominent band in Figure~\ref{Fig3} is the $Q$-branch of
CO$_2$ at 15 $\mu$m. Its $R-$ and $P-$branch lines are weaker but
still visible, most notably in IRAS 23385. The $Q-$branches of the
CO$_2$ hot bands and of the 13-C isotopolog of CO$_2$ are also
detected in some sources, most notably GW Lup \cite{Grant23}. An
irregular set of lines due to H$_2$O shows up most strongly in the Sz
98 disk, but these lines are also present in the GW Lup and V1094 Sco
disks. In contrast, in the J160532 disk only a few weak H$_2$O lines
are tentatively seen \cite{Tabone23}, and none in IRAS 23385 in this
wavelength range. The H$_2$ S(1) line at 17.0 $\mu$m and H I
recombination lines are also seen in a few sources, as are some OH
lines. Excitingly, bands of C$_6$H$_6$ and C$_4$H$_2$ are clearly
detected in the J160532 disk (Figure~\ref{Fig4}), and CH$_4$ emission
is possibly found as well \cite{Tabone23}. Many of these lines are seen
with JWST for the first time in disks.

Note that the bulk of the lines in this wavelength range can be fit
with these molecules. There could be additional lower abundance
molecules present, or isotopologs of identified molecules, whose
weaker lines are blended with the stronger lines and that have not yet
been identified at this stage of data processing. Also, line lists of
various potentially interesting molecules are currently unavailable or
incomplete at high temperatures, especially for hydrocarbons (see also
paper by Kamp et al.\ in this volume).

\begin{table*}[h]
\small
  \caption{Best fit molecular model parameters}
  \label{Table1}
  \begin{tabular*}{\textwidth}{@{\extracolsep{\fill}}lrccrcc}
    \hline
    Molecule &  & IRAS 23385 & & & GW Lup   \\
             &  $N$ & $T$ & $R$ &  $N$ & $T$ & $R$  \\
             & (cm$^{-2}$) & (K) & (au) & (cm$^{-2}$) & (K) & (au) \\
    \hline
    CO$_2$     &  1E17 & 150 & 35 & 1E18 & 475 & 0.09  \\
    H$_2$O     &  $<$1E18 & [150] & [35] & 2E18 & 625 & 0.16 \\
    C$_2$H$_2$ &  7E16& 210 & 8 & 4E17 & 550 & 0.05 \\
    \hline
  \end{tabular*}
Note: brackets indicate parameters that were fixed in the fit.
\end{table*}

\subsection{Inferred column densities and temperatures}

\begin{figure}[tbh]
\centering
  \includegraphics[width=0.48\textwidth]{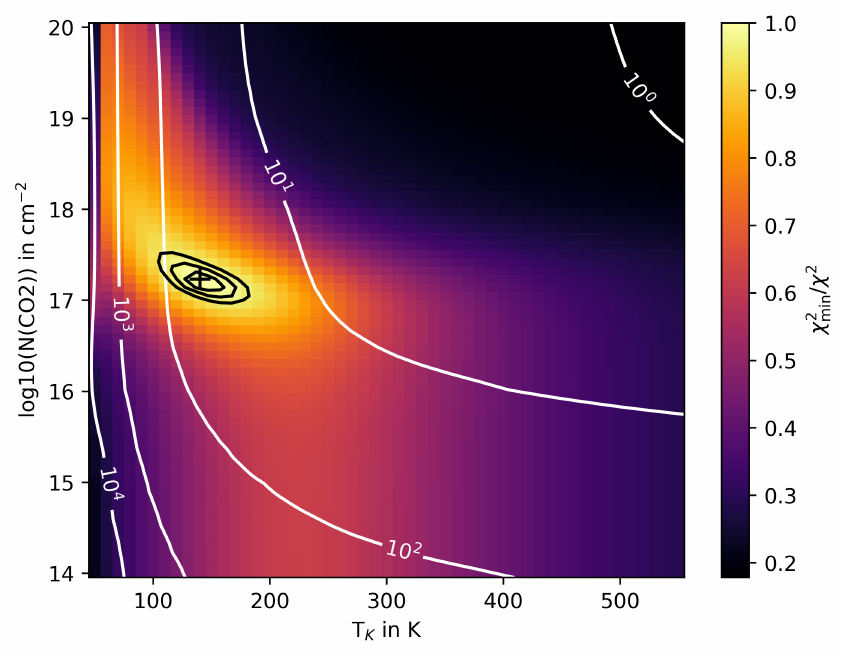}
  \caption{Example of a $\chi^2$ map to constrain temperature and
    column density, here for the case of CO$_2$ in IRAS 23385. The
    white lines indicate the best fitting emitting areas for those
    $N$, $T$ combinations. The black lines indicate the $1\sigma$,
    2$\sigma$ and 3$\sigma$ contours.}
  \label{Fig5}
\end{figure}

Table~\ref{Table1} summarizes the inferred best-fitting column
densities, temperatures and emitting areas of various molecules for
two of the sources. An example of a typical $\chi^2$ plot can be found
in Figure~\ref{Fig5}. As found in previous studies based on {\it
  Spitzer} data \cite{Carr11,Salyk11}, there is a wide range of fit
parameters that reproduce the spectra well. Most notably, in the low
column density regime (typically $<10^{17}$ cm$^{-2}$ depending on
molecule) where lines are optically thin, there is a complete
degeneracy between column density and emitting radius. On the other
hand, the latter can be determined from optically thick lines whose
strength scales directly with emitting area. The temperature is often
well constrained from the shape and position of the $Q-$branch
feature, where lines with $\Delta J$=0 pile up: the warmer the gas,
the broader the feature. MIRI can resolve this $Q-$branch much better
than before. An excellent example of the latter is provided by the
CO$_2$ $Q-$branch toward IRAS 23385, which is only consistent with a
low temperature of $150 \pm 50$ K. The availability of both optically
thick and thin lines with JWST as well as their 13-C isotopologs
helps to resolve several of these degeneracies.

A few column density ratios are summarized in Table~2 to give an
impression of the range of values, which will be discussed further below. For
V1094 Sco, these column density ratios are obtained from a preliminary
analysis assuming the same emitting radius of 0.13 au for all
species. For more details, see refs.\cite{Grant23,Tabone23,Francis23}.

\section{Discussion}

\subsection{Diversity in spectra}

Figures 1--4 illustrate the diversity in mid-infrared spectra of
embedded protostars and young stars with disks, not just between these
two classes of sources but even between disk-dominated sources
themselves. High quality mid-infrared spectra are now possible for
much fainter sources than before, down to mJy continuum level, thanks
to JWST. Focusing just on the molecular emission, one of the striking
features of Figure~\ref{Fig3} is the difference in the strength of
H$_2$O emission in the 14.3--17.3 $\mu$m region, with H$_2$O lines
virtually absent in the spectra of the high mass binary IRAS 23385 and
the very low-mass source J160532, but clearly present in those
of GW Lup and V1094 Sco and dominant in Sz 98. Another obvious
difference is the strength of the C$_2$H$_2$ band with respect to
other features, with C$_2$H$_2$ and hydrocarbons dominating the
spectrum of the very low-mass star J160532 but being much weaker
toward other sources (see paper by Kamp et al., this volume). CO$_2$
is detected in all sources, most strongly relative to other features
in the GW Lup disk.

How about nitrogen? One molecule that has not yet been detected in
disks with MIRI is NH$_3$ whose $\nu_6$ mode at 8.8 $\mu$m can now be
observed at much higher spectral resolution with the MRS than was
possible with {\it Spitzer}.  The HCN $Q-$branch at 14.0 $\mu$m is
seen in most sources but is now recognized to be heavily blended with
both C$_2$H$_2$, CO$_2$ and/or H$_2$O lines (Figure~\ref{Fig4})
resulting in an HCN contribution that is smaller than thought based on
{\it Spitzer} data \cite{Salyk11,Najita13}.  This means that an even
lower fraction of nitrogen has been identified than before, suggestion
that the bulk of the nitrogen in the inner disk may well be in the
unobservable N$_2$ form \cite{Pontoppidan19}.

Where does the molecular emission originate? For the low-mass sources,
the lines clearly come from the inner region of their disks, typically
within 1 au, as is also evidenced by their small emitting areas
(Table~1). Although the emission could in principle also come
  from a narrow ring further out in the disks, the high inferred
  temperatures also point to a location in the inner disk.

  For high-mass protostars, the emitting area of only 35 au radius is
  also relatively small.  Interestingly, for high-mass protostars
  these molecular bands are often seen in absorption rather than
  emission in spectra taken with the {\it Infrared Space Observatory}
  \cite{Helmich96,Lahuis00,Boonman03CO2}. Those lines were thought to
  originate in the inner warm envelope where ices have sublimated
  (``hot core'') or in shocked gas along the line of sight seen in
  absorption toward the strong mid-infrared continuum due to hot dust
  close to the protostars. However, more recent high spectral
  resolution ground-based and SOFIA-EXES data suggest an origin in the
  inner part of a disk around the massive young star
  \cite{Knez09,Barr20}. Weak water absorption at 6 $\mu$m is seen
  toward IRAS 23385 but only towards source B of the binary
  \cite{Francis23}. Emission lines of CO$_2$, C$_2$H$_2$ and HCN at
  13--15 $\mu$m have previously been seen only toward the Orion peak 1
  and 2 shock positions away from the strong continuum
  \cite{Boonman03Orion} and in the Cepheus A high-mass region
  \cite{Sonnentrucker06,Sonnentrucker07}. Interestingly, in both cases
  moderate excitation temperatures of $\sim$200 K were found, lower
  than those seen for absorption bands, and similar to those derived
  here. For IRAS 23385, the molecular emission is clearly centered on
  the two protostars and is not detected in off-source shock positions
  \cite{Francis23}. Also, the lack of strong H$_2$O emission,
  universally observed to be associated with shocks
  \cite{vanDishoeck21}, argues against an outflow origin. Although we
  cannot exclude a shock origin at this stage, we assume here that the
  emission originates in the disk(s) around (at least one of) the
  protostars in the IRAS 23385 system.

\begin{table}[h]
\small
  \caption{Column density ratios of various molecules}
  \label{Table2}
  \begin{tabular*}{0.48\textwidth}{@{\extracolsep{\fill}}lrrrr}
    \hline
    Ratios & IRAS23385 & V1094Sco & GW Lup & J160532 \\
    \hline
    CO$_2$/H$_2$O     & $>$0.1 & 0.004 & 0.45 & $>$0.1 \\
    CO$_2$/C$_2$H$_2$ & 1.7 & 0.1 & 2.2 & 0.0001 \\
    \hline
  \end{tabular*}
Note: column density ratios do not reflect local abundance ratios. For J160532, the optically thick C$_2$H$_2$ component is used to compute the ratio.
\end{table}

\subsection{Diversity in chemistry}

Table~2 summarizes the column density ratios of the main molecules
considered here in four of the sources. Uncertainties are up to an
order of magnitude, based on the $\chi^2$ plots. It should be stressed
that these ratios should not be equated with abundance ratios since
the emission of different molecules (or even of different bands of the
same molecule) may originate from different regions or layers of the
disk \cite{Bruderer15,Woitke18}. Moreover, the emission seen at
mid-infrared wavelengths only probes the upper layers of the disk
above the $\tau_{\rm mid-IR}=1$ contour where the dust continuum
becomes optically thick.

In spite of these considerations, the similarities and differences in
mid-infrared spectra can give some insight into the chemical processes
at play in planet-forming zones of disks, especially since the
differences in column density ratios are much larger than those of
flux ratios of optically thick lines. As concluded from the appearance
of the spectra, the J160532 inner disk is particularly rich in
C$_2$H$_2$, whereas the V1094 Sco has the lowest CO$_2$/H$_2$O ratios, with
Sz 98 likely even lower \cite{Gasman23}. Apart from the larger size of
the emitting region, the high-mass source IRAS 23385 does not stand
out in terms of its abundance ratios compared with the orders of
magnitude lower luminosity T Tauri sources.

GW Lup is part of a handful of sources that were found by {\it
  Spitzer} to have strong CO$_2$ emission but no emission from other
molecules, the so-called ``CO$_2$-only'' sources
\cite{Salyk11}. Thanks to the detection of $^{12}$CO$_2$ hot bands,
its $P-$ and $R-$branch lines and the $^{13}$CO$_2$ $Q-$branch, the
inferred CO$_2$ column density is two orders of magnitude higher (and
emitting area smaller) than found in analyses of the {\it Spitzer}
data, illustrating the importance of the additional weaker optically
thin features now observed with MIRI-MRS \cite{Grant23}.  H$_2$O and
other molecules are now weakly detected as well with JWST, but its
CO$_2$/H$_2$O column density ratio remains clearly higher than that of
other low-mass sources.

\subsection{Chemical processes}

\subsubsection{Disk models.}

The chemistry in disks is known to vary both radially and vertically
owing to large gradients in temperature, density and UV irradiation. A
detailed 2D (or axisymmetric 3D) model of the physical structure of a
disk therefore needs to be constructed before the chemistry can be
addressed.  Such thermochemical models have been developed for disks
around low-mass stars by various groups
\cite{vanDishoeck06,Woitke09,Bruderer12,Walsh12}, with some of them 
focused specifically on inner disk chemistry where high-temperature
gas-phase chemistry dominates
\cite{Agundez08,Woods09,Glassgold09,Walsh15,Woitke18,Anderson21}.  Snowlines,
i.e., the location in the disk where 50\% of the molecule is in the
gas and 50\% in the ice, play an important role in setting the overall
gas-phase chemical composition. Of particular interest is the C/O
ratio in volatiles (i.e., gas + ices), since ices naturally lock up
more oxygen than carbon \cite{Oberg11co}. For CO, whose pure ice
binding energy is $E/k \approx$855 K \cite{Bisschop06}, its snowline
lies around 20 K which is typically at tens of au in a disk around a
low-mass star \cite{Long18}. For H$_2$O, whose binding energy is much
higher at $E/k\approx$5600 K \cite{Fraser01}, its snowline is at
$\sim$160 K at the high densities in disk midplanes, located at less than a
few au \cite{Mulders15}. The snowlines of other molecules like CO$_2$
generally lie in between those of CO and H$_2$O \cite{Minissale22},
with the exception of those of large organic molecules. In particular,
the ``soot'' line due to sublimation and erosion of refractory hydrocarbon
material is thought to lie around 500 K in disks \cite{Li21}.

If $R$ listed in Table~1 is taken to be a disk radius, the emission
observed here for low-mass stars originates from well inside the
H$_2$O snowline. At these radii and temperatures, much of the
oxygen that was locked up in ices should have returned to the gas
phase \cite{Oberg11co}. The overall (gas + ices + refractory dust) C/O
ratio in disks is expected to be similar to the stellar value, which
in turn should be close to the solar value of C/O=0.59 within the
solar neighborhood \cite{Asplund21}. For gas-phase chemistry outside
the refractory dust sublimation radius, only the volatile (gas + ice)
C/O ratio is relevant, which is generally taken to be somewhat lower
than 0.6 in disk models. Its precise value does not matter much for
the chemistry and mid-infrared line emission
\cite{Woitke18,Anderson21} as long as C/O$<$1.

It is important to note that any disk model reproducing mid-infrared
lines needs to have a gas/dust ratio in the upper layers that is
significantly higher than the standard interstellar medium value of
100, typically \cite{Meijerink09,Bruderer15,Woitke18} of order
$10^3-10^4$. Such high gas/dust ratios are naturally explained by
grain growth and settling to the midplane \cite{Greenwood19}.

The physical and chemical structure of disks around high-mass O and B
stars has not been modeled in as much detail as that of disks around
A- and later-type stars \cite{Walsh15}. An additional complication for
such high-mass stars is that they are much younger and do not
have an optically visible pre-main sequence stage. Although disks are
commonly found to form in simulations of high-mass star formation
\cite{Kuiper11,Zhao20}, they are still embedded in their natal
envelopes.  Moreover, accretion rates are still high, so heating of
their inner disk midplane is dominated by viscous processes rather
than passive irradiation \cite{DAlessio98}. Figure~\ref{Fig7} compares
the 2D dust temperature structure for one quadrant of a disk around a
low-mass star with that of a high-mass star
\cite{Nazari22LM,Nazari22HM}, obtained using the RADMC-3D code
\cite{Dullemond12}. The gas temperature is closely coupled to that of
the dust except in the upper surface layers. For high-mass disks, the
midplane is actually warmer than the surface layers within the inner
$\sim$100 au \cite{Nazari22HM} (depending on the accretion rate)
giving rise to infrared absorption lines \cite{Barr20}. At larger
radii, the disk switches to the usual structure of a vertically
decreasing temperature structure from surface to midplane leading to
mid-infrared emission lines.  IRAS 23385 has a somewhat lower
luminosity and accretion rate than assumed in the model in
Figure~\ref{Fig7}, suggesting that the switch moves well inside 100
au.

\subsubsection{CO$_2$ versus H$_2$O: cavities and dust traps.}

Armed with the above background information on disk models, we now
address the chemistry of the main species considered in this paper.

\paragraph{Chemical processes relating CO$_2$ and H$_2$O.}
The temperature structure of the inner disk plays a crucial role in
setting the balance between the CO$_2$ and H$_2$O abundances
\cite{Walsh15,Bosman22CO2}. Both molecules are formed through
gas-phase reactions with the OH radical \cite{vanDishoeck13,Walsh15}.
At temperatures of 100--250 K, the OH + CO $\to$ CO$_2$ + H reaction
produces CO$_2$ with typical abundances of $10^{-7}-10^{-6}$ with
respect to total H. At higher temperatures, the reaction OH + H$_2$
$\to$ H$_2$O + H, which has an energy barrier of 1740 K
\cite{Baulch92}, takes over and produces H$_2$O in favor of CO$_2$.
Once H$_2$O becomes abundant, its column density becomes high enough
($\sim 10^{18}$ cm$^{-2}$) to become self-shielding \cite{Bethell09}
pushing the balance even further toward H$_2$O. Note that OH itself
also needs moderately warm temperatures to form since the O +
H$_2$ $\to$ OH + H reaction is endothermic by 900 K and has an energy
barrier as well \cite{Baulch92}. Both the formation of
OH and H$_2$O are boosted by the presence of vibrationally excited
H$_2$ where the vibrational energy can be used to overcome the
energy barriers.

H$_2$O and CO$_2$ are both abundant components of ices in cold
interstellar clouds and protostellar envelopes \cite{Boogert15}. A
fraction of these ices are likely preserved in the transition from
envelope to disk, especially for the stronger bound molecules like H$_2$O
\cite{Visser09}, as also illustrated by the similarity between
interstellar and cometary ice abundances
\cite{Bockelee00,Drozdovskaya19}. Inside their respective snowlines,
the ice species sublimate and thus add to the gas-phase chemistry
budget. Minor species intimately mixed with H$_2$O ice sublimate
together with water at its snowline \cite{Collings04}.

\begin{figure*}[t]
\centering
\vspace{-1.5cm}
  \includegraphics[width=0.8\textwidth]{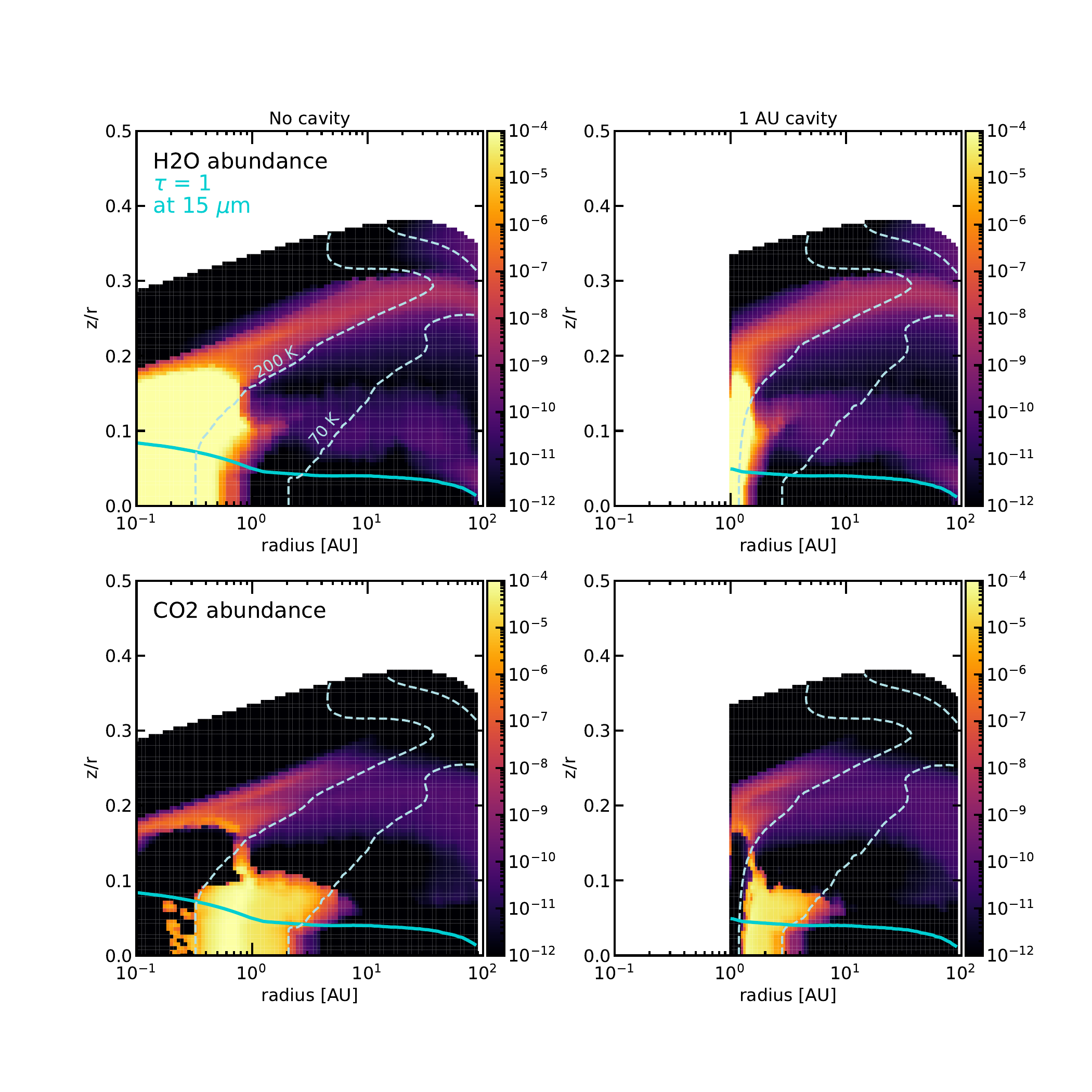}
\vspace{-1cm}
\caption{Two dimensional map of the H$_2$O and CO$_2$ abundances of
  one quadrant of a disk around a low-mass star without a cavity
  (left) and with a 1 au cavity (right) \cite{Vlasblom23}. The dashed
  lines indicate the 70 and 200 K temperature contours; the blue line
  the $\tau_{\rm mid-IR}=1$ line at 15 $\mu$m. The cavity radius is
  located between the H$_2$O and CO$_2$ snowlines. The disk model
  parameters are taken from ref.\cite{Bosman22CO2}.}
      \label{Fig6}
\end{figure*}

Bosman et al.\cite{Bosman22H2O,Bosman22CO2} find that H$_2$O
mid-infrared emission dominates over CO$_2$ emission in the 13--17
$\mu$m range for low-mass disks, if H$_2$O self-shielding and
efficient gas heating mechanisms are included. In particular, so-called
(photo)chemical heating, in which the excess energy released by
photodissociation and subsequent chemical reactions of the
dissociation products is used to heat the gas \cite{Glassgold15}, can
raise the temperatures in the upper layers significantly
\cite{Adamkovics14}.

\paragraph{Role of small cavities.}
Given the model predictions that H$_2$O emission should dominate
mid-infrared spectra, which processes could then make CO$_2$ more
prominent than H$_2$O in some disks but not others? One parameter that
affects mid-infrared line fluxes is the amount of small grains in the
upper layers. However, both CO$_2$ and H$_2$O fluxes are decreased by
comparable amounts if the gas/dust ratio is decreased
\cite{Antonellini15}.  Another option would be to change the C/O
ratio, but again both CO$_2$ and H$_2$O emission lines appear to be
affected similarly \cite{Woitke18,Anderson21}. Walsh et
  al.\cite{Walsh15} show that for very low-mass stars H$_2$O formation
  is indeed suppressed in the disk's surface layers due to their lower
  temperatures, consistent the lowest-luminosity sources in our
  sample, GW Lup and J160532, showing the weakest H$_2$O
  lines. However, some higher luminosity sources, including IRAS
  23385, also show weak water emission.

An alternative explanation would be to introduce a deep dust +
gas cavity in the inner disk, caused for example by a companion. Its
radius needs to be in between that of the H$_2$O and CO$_2$ snowlines
in order to remove most of the H$_2$O but not the CO$_2$ emission
\cite{Grant23,Antonellini16,Anderson21}. Figure~\ref{Fig6} presents an
example of the 2D abundance structure of H$_2$O and CO$_2$ in a model
\cite{Vlasblom23} obtained using the DALI thermochemical
code\cite{Bruderer12} for a disk with the same parameters as in Bosman
et al.\cite{Bosman22CO2}. Two cases are presented: a full disk and a
disk with a deep dust + gas cavity out to 1 au, a location that is in
between the two snowlines. The 70 and 200 K contours are also included
to highlight the temperature range where CO$_2$ is efficiently
produced.

If there is no cavity, it is seen that H$_2$O is highly abundant in
the inner disk above the $\tau_{\rm mid-IR}=1$ (where the 15 $\mu$m
dust continuum becomes optically thick) and would thus produce very
strong H$_2$O lines. However, its column relative to CO$_2$ is
strongly reduced in the model with a cavity (except at the very edge
of the cavity). These models therefore provide a proof of concept
for explaining those sources with high CO$_2$/H$_2$O column density
ratios with disks that have small cavities.

For the GW Lup disk, the H$_2$O and CO$_2$ snowlines are expected to
lie around 0.4 and 1.4 au, respectively \cite{Temmink23}, so a
qualitatively similar effect can be expected. Such cavities are too
small to be resolved with ALMA but could be revealed in their spectral
energy distributions or in spectrally resolved CO ro-vibrational line
profiles \cite{Salyk09,Pontoppidan08,Antonellini20}. With a few
exceptions, these low-mass stars and their disks are generally too
weak to image with infrared interferometers \cite{Gravity21Perraut}.

\begin{figure}[tbh]
\centering
  \includegraphics[width=0.44\textwidth]{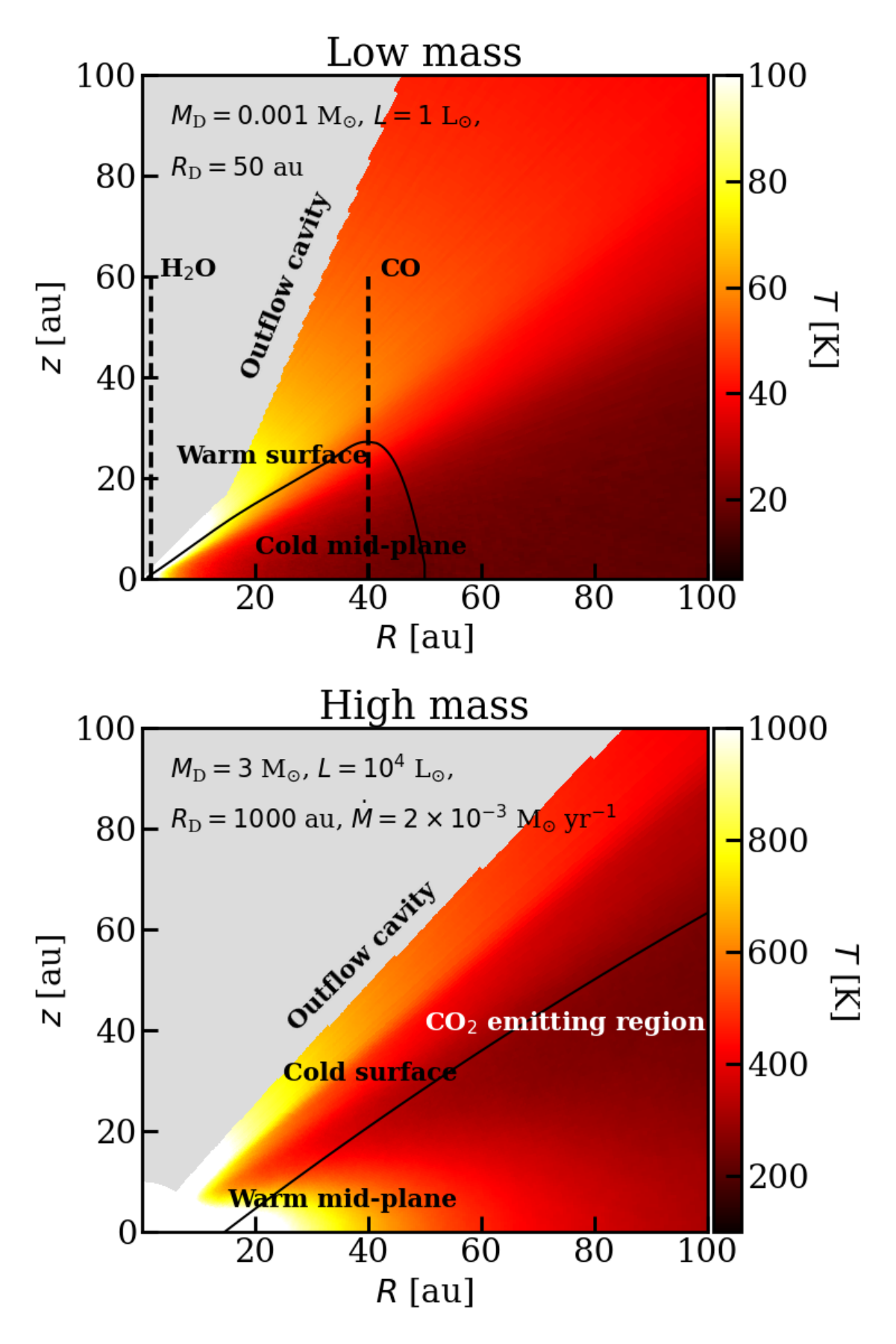}
  \caption{Two dimensional map of the dust temperature structure of a
    disk around a low-mass star (top)\cite{Nazari22LM} with that
    around a high-mass star (bottom)\cite{Nazari22HM}.  The solid
    black line shows the approximate location of the disk, which is
    still embedded in an envelope for the high-mass source.  Note that
    for the high-mass disk, the midplane temperature in the inner
    region ($\leq$50--60 au) is larger than that at the disk surface
    due to accretional heating; no such heating is included for the
    low-mass source. Such a structure gives rise to mid-infrared
    absorption lines. At larger radii, the surface is warmer than the
    midplane due to irradiation resulting in mid-infrared emission
    lines (indicated with CO$_2$ emitting region). Model parameters
    \cite{Nazari22LM,Nazari22HM} are for low mass: $M_*$=0.5
    M$_\odot$, $L= 1$ L$_\odot$, $M_{\rm disk}$=0.001 M$_{\odot}$;
    high mass: $M_*$=30 M$_\odot$, $L= 10^4$ L$_\odot$,
    $M_{\rm disk}$=3 M$_{\odot}$, ${\dot M}=2\times 10^{-3}$ M$_\odot$
    yr$^{-1}$. Both models use dust opacities appropriate for large
    grains \cite{Nazari22LM}. For the low-mass disk, the H$_2$O
    snowline is at 1.5 au and the CO snowline at 40 au; for the
    high-mass disk, they are at 390 au and $>$20000 au, respectively.}
  \label{Fig7}
\end{figure}

Interestingly, the strong CO$_2$ emission and lack of H$_2$O in the
high-mass source IRAS 23385 also fits into this scenario. The only
difference is that there is not necessarily a cavity but rather a
temperature inversion in the inner disk which would cause the
H$_2$O-rich gas to be in absorption (most notably in its $\nu_2$ band
at 6 $\mu$m \cite{Francis23}) rather than emission. Figure~\ref{Fig7}
shows that the temperature in the surface layers in the region
where accretional heating no longer dominates (and where lines go
back to being in emission), is around 200 K. Such a relatively cool
temperature is consistent with that found for the CO$_2$ emission in this
source (Table~1).

\paragraph{Role of dust traps.}
For low-mass sources, a related effect could be the existence of a
dust trap triggered by a pressure bump due to a companion that created
the cavity. If that dust trap lies between the H$_2$O and CO$_2$
snowlines, it locks up water-rich pebbles thereby suppressing
H$_2$O. However, some gas enriched in carbon
\cite{Miotello17,Cleeves21,Bosman21MAPSVII}, or in CO$_2$ due to 
ice sublimation inside its snowline, could still be transported through
the trap to the inner disk.

\begin{figure*}[tbh]
\centering
  \includegraphics[width=0.7\textwidth]{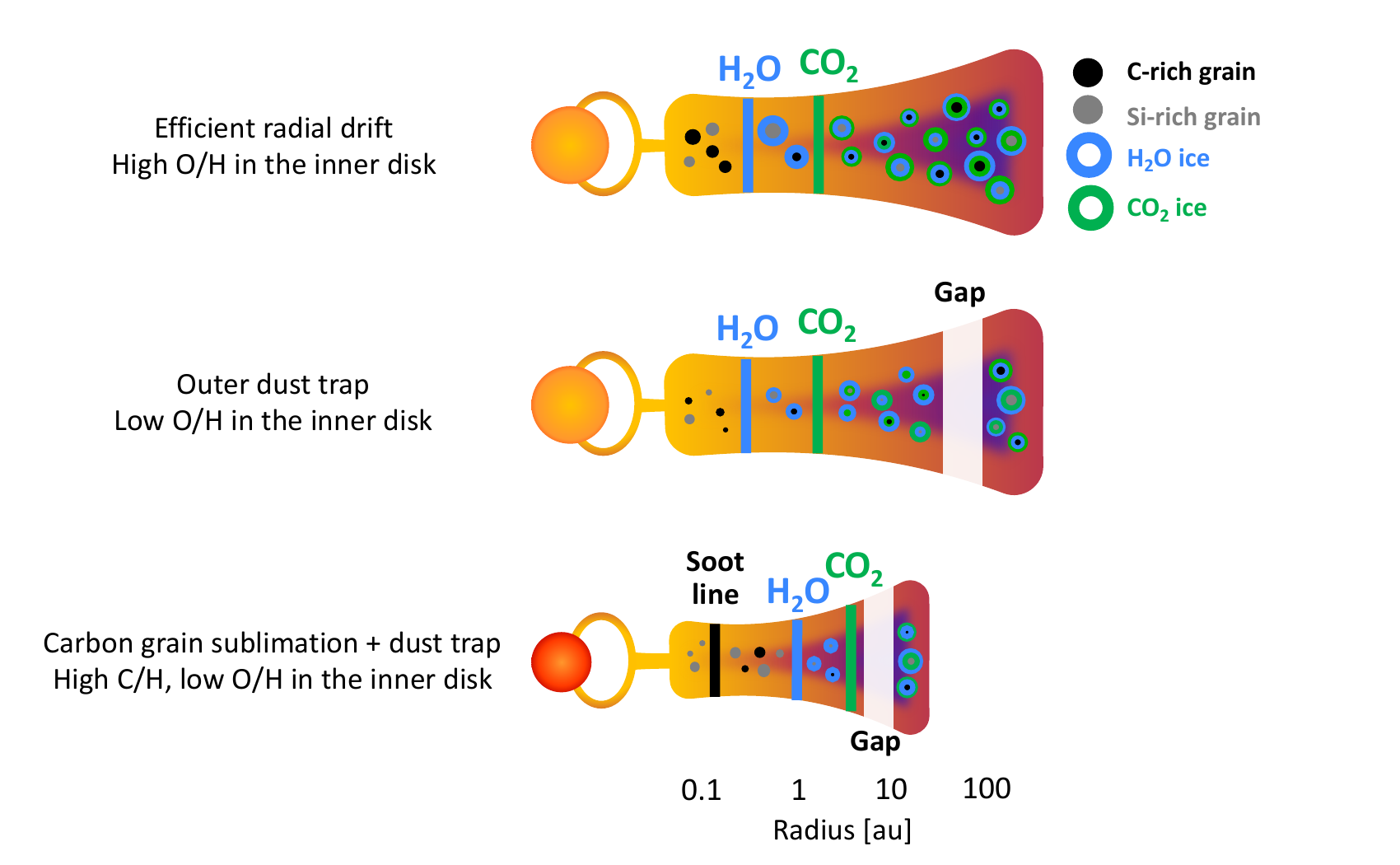}
  \caption{Cartoons illustrating the cases of (a) enhanced H$_2$O and
    O/H in the inner disk due to icy pebble drift; (b) reduced H$_2$O
    and O/H in the inner disk due to a dust trap outside the H$_2$O
    snowline. If this trap is still inside the CO$_2$ snowline, CO$_2$
    could be enhanced; and (c) as (b) but for a very low-mass star
    with enhanced C$_2$H$_2$ at the soot line.}
  \label{Fig8}
\end{figure*}

A dust trap locking up the bulk of the H$_2$O ice would also lower the
overall O/H abundance ratio in the inner disk \cite{Banzatti20}. The
same holds for C/H, if CO, CH$_4$ and/or carbon-containing molecules
are trapped. Some inner disks are indeed known to have low O/H and/or
C/H abundances \cite{Bosman19TWHya,McClure19,Sturm22}. Disks that show
spatially resolved dust traps with ALMA \cite{Andrews18,Long18} are
likely in this category, provided that the dust traps formed early in
the disk's evolution, before the bulk of the grains have moved inwards.

In contrast, any disk in which drifting icy pebbles from the outer
disk have reached the inner disk and crossed the H$_2$O snowline
unimpeded should have O/H abundances that could be enhanced by 1--2
orders of magnitude \cite{RBooth19,Bosman18CO2}. Similarly, CO can be
increased in the inner disk due to drifting icy grains
\cite{Booth19,Zhang19,Krijt20}.  Since these icy pebbles are also
thought to contain considerable amounts of CO$_2$ ice, this should not
change the relative ratios of the H$_2$O and CO$_2$ lines if both are
continuously being supplied to the inner disk \cite{Bosman18CO2},
unless the extra CO$_2$ would remain hidden below the
$\tau_{\rm mid-IR}=1$ line due to a lack of vertical
  mixing. Disks which show compact dust continuum emission with ALMA
are good candidates to test this pebble drift scenario.

Figure~\ref{Fig8} illustrates these possibilities in a cartoon. It is
clear that the combination of large samples of disks studied with both
JWST mid-infrared spectroscopy and with ALMA at its highest spatial
resolution, including disks that are small and currently unresolved
with ALMA at moderate spatial resolution, can provide insight into the
relative importance of these processes.

\subsubsection{Rich hydrocarbon chemistry: ``soot'' line and C/O$>$1.}

The J160532 disk stands out in having unusually strong bands of
C$_2$H$_2$ and other hydrocarbons including species that have not been
identified in disks before the advent of JWST \cite{Tabone23}. Equally
interesting is the discovery\cite{Tabone23} that the broad bands at
7.7 and 13.7 $\mu$m are due to very high column densities of hot
C$_2$H$_2$.

What could be causing such abundant C$_2$H$_2$ in this disk?  It is
known that PAHs and carbonaceous grains can be destroyed by UV
radiation, by chemical processes (e.g., reactions with H, O or OH) or
by grain sublimation under the conditions in the inner disk, with
C$_2$H$_2$ as a main product \cite{Kress10,Anderson17}. When and where
the carbon grain sublimation and erosion takes place depends on the
type of carbonaceous material. Some laboratory data put the
sublimation of amorphous carbon grains at 1200 K \cite{Gail17} whereas
other experiments suggest that thermal decomposition of refractory
hydrocarbon material (also known as kerogen-like insoluable organic
matter, IOM) occurs at lower temperatures, around 400-500 K depending
on heating timescale \cite{Chyba90,Kebukawa10,Li21}. This IOM material
is thought to make up the bulk of the refractory carbon. Aliphatic and
aromatic hydrocarbons, which contain of order 10\% of total carbon,
sublimate at somewhat lower temperatures of 300--400 K
\cite{Gail17}. Together, they form the so-called hydrocarbon “soot”
line around 400--500 K, a term coined by Li et al.\cite{Li21} to
indicate the location where the destruction of hydrocarbon solids
occurs. This temperature matches well with that found for the
high abundance C$_2$H$_2$ gas component. One possibility is therefore
that hydrocarbon grain destruction is being observed directly in this
disk.

The destruction of carbon grains by thermal processing is thought to
be irreversible.  On the other hand, the conditions of warm dense gas
with high C$_2$H$_2$ abundances are conducive to efficient formation
of small aromatic molecules and PAHs up to temperatures where
degradation starts to take over \cite{Frenklach89,Morgan91}. Indeed,
the production of the smallest aromatic molecule, benzene, in the
inner regions of disks as observed here was predicted by Woods \&
Willacy \cite{Woods07,Woods09}. The J160532 and other similar disks
around very low mass stars \cite{Arabhavi23} are clearly excellent
laboratories for studying such hydrocarbon chemistry in detail.

A related interesting question is how common the broad 7.7 and 13.7
$\mu$m features indicative of a ``soot'' line are in other
sources. While disks around very low-mass stars are known to have
strong C$_2$H$_2$ bands based on {\it Spitzer} data
\cite{Dahm09,Pascucci09,Pascucci13}, J160532 is the only such disk so far in
which the broad bumps are clearly seen. What could make the J160532
disk special? One option is that we are able to look deep into the
J160532 disk close to the midplane due to the lack of small grains, as
evidenced by the large gas columns and strong H$_2$ lines
\cite{Tabone23}. In other disks, the midplane ``soot'' line could be
hidden below the $\tau_{\rm mid-IR}=1$ line. Another option could be
that we are witnessing the J160532 disk at a special time when
destruction is taking place, perhaps triggered by a recent heating
event due to accretion.

The J160532 disk presents another puzzle, namely that of water. The
disk is clearly warm enough to sublimate any water ice from the grains
and to trigger rapid formation of H$_2$O through the reactions of O
and OH with H$_2$. In disk chemical models, such high abundances of
C$_2$H$_2$ compared with H$_2$O (Table~2) are only found if the
volatile C/O ratio of the gas in the inner disk is significantly
increased compared with standard values of volatile
C/O=0.4--0.5. Specifically, C/O ratios $>$1 are needed to boost the
hydrocarbon production and lock up the bulk of the volatile oxygen in
CO and some CO$_2$ \cite{Najita13,Woitke18,Anderson21}. A small amount
of H$_2$O could still be produced in the upper layers of the disk
where destruction of CO by UV photons or by He$^+$ frees up some
atomic O that can react with H$_2$ to form H$_2$O.  Walsh et
  al.\cite{Walsh15} note that the atmospheres of disks around very low
  mass stars may naturally {\it appear} to be carbon rich (without the
  need for a high C/O ratio) because the unobservable O$_2$ rather
  than H$_2$O becomes the main gaseous oxygen carrier. However, the
  orders of magnitude higher C$_2$H$_2$ column and C$_2$H$_2$/CO$_2$
  ratio observed for J160532 than found in these models strongly
  points to a genuine carbon-rich atmosphere with C/O$>$1.

The most plausible mechanism to lead to C/O$>$1 and prevent oxygen
from entering the inner disk would be to lock most of it up in water
ice in pebbles and planetesimals in the outer disk beyond the water
snowline, preventing them from migrating to the inner disk
\cite{Najita13,Pascucci13}, as illustrated in Figure~\ref{Fig8}
(bottom).

\subsubsection{Implications for planet formation}

The JWST data provide detailed insight into the chemical composition
of the gas in the upper layers of the disk.  Planetesimals and planets
are, however, formed in the disk midplanes which are not probed
directly by mid-infrared data. Nevertheless, indirect information can
be obtained if mid-plane material is vertically mixed upward. For
example if icy pebbles drift across the water snowline and enhance
the gaseous O/H ratio of the midplane gas, this can be mixed upward on
comparatively short timescales (e.g.,
refs.\cite{Semenov11,Krijt20,Woitke22}) to result in enhanced OH and
H$_2$O emission.

There is also downward gas transport: vertical transport of molecules
can be followed by freeze-out onto the colder large grains in the
midplane. If those grains become too large to be lifted back up, the
disk surface composition becomes part of the planetesimal building
material. This so-called ``vertical cold finger effect'' also limits
the radial extent of the molecular emission: snow surfaces become
vertical walls at the midplane snow radius rather than being curved
with height in the disk, an effect that is well known for H$_2$O
\cite{Meijerink09,Blevins16,Bosman21H2O} but likely also holds
for other molecules.  Finally, if a gap has opened in a disk, surface
layer gas can be accreted directly onto forming giant planets through
meridional flows \cite{Morbidelli14,Teague19} thus leaving an imprint
on the composition of their atmospheres.

Whether or not carbon-rich planets will be built in the J160532 disk
is still an open question and depends also on the time of planet
formation. A significant fraction of the carbon is in the gas phase
and refractory carbon grains may even have been destroyed in this
disk. If this carbon-rich gas were dispersed quickly from the system, only a small
fraction of carbon would eventually be included in planets. This
scenario has been proposed to explain why Earth is so carbon poor
\cite{Lee10,Li21}.

\section{Conclusions}

The early JWST MIRI-MRS spectra presented here provide examples of how
its increased spectral resolution and sensitivity provide new insights
into the chemical composition of disks, from young to mature disks and
from low- to high-mass stars.  In particular, the detection of new
species like C$_4$H$_2$ and C$_6$H$_6$ allows proposed chemical
pathways to be tested. New lines of known species and of their
isotopologs, with first detections of $^{13}$CO$_2$ and
$^{13}$C$^{12}$CH$_2$, help to break degeneracies between model fit
parameters so that their temperatures and column densities can be
constrained more accurately. Indeed, orders of magnitude differences
in inferred column densities have been found compared with earlier
work. The lack of NH$_3$ suggests that most of the volatile nitrogen
is in unobservable species.

The data highlight the diversity in mid-infrared spectroscopy of disks
hinted at by {\it Spitzer}, from sources dominated by CO$_2$
with almost no H$_2$O, to those rich in H$_2$O to disks showing
booming features of C$_2$H$_2$. There are also some similarities: the
molecular emission observed in the high-mass protostellar source IRAS
23385, thought to originate from a massive young disk, indicates
similar abundance ratios as those found in some CO$_2$-rich low-mass
disks.

The diversity in chemistry and spectra is likely related to
differences in the physical structures of the disks. Lines are
brighter in disks around more luminous sources, which push the
emitting area to larger radii and thus boost optically thick
lines. The lack or weakness of lines of certain molecules, most
notably H$_2$O, can be due to the very low luminosity of some
  sources in our sample. However, it is likely also linked to the
presence of small cavities in the inner disk that cannot be resolved
with ALMA and whose positions extend beyond the water snowline. A
related possibility is the presence of dust traps outside of the
respective snowlines of different molecules that can lock up certain
elements, most notably oxygen, and prevent it from entering the inner
disk.

The JWST data also provide evidence for a phenomenon that has been
postulated but not seen prior to JWST: the destruction of hydrocarbon
grains at the ``soot'' line boosting the C$_2$H$_2$ abundance by
orders of magnitude\cite{Tabone23} and raising the C/O ratio to $>$1.

There are a number of obvious next steps in this research. First,
larger samples of disks have to be studied with JWST MIRI-MRS to
search for more similarities and probe the limits of diversity, from
young to old ``debris'' disks and across the stellar mass range. MRS
spectra of line-rich sources will allow to hunt for other minor
species, including NH$_3$. Also, adding NIRSPEC-IFU data for the same
sources will be highly valuable to cover the CO ro-vibrational bands
at 4.7 $\mu$m as well as the stretching bands of most of the species
discussed here. Molecular data on vibration-rotation lines of many
molecules, including their isotopologs, are however still incomplete
and more laboratory work is needed.

Second, ALMA observations at the highest spatial resolution (down to a
few au) are needed to search for small-scale substructure in the
continuum that are indicative of dust traps. Their location with
respect to different snowlines will help to understand which elements
are locked in the outer disk and unable to reach the inner disk. In
contrast, some disks may be dominated by drifting icy pebbles that
enhance the inner disk in heavy elements. More work is also
  needed on high resolution studies of the physical structure of
  high-mass disks with ALMA.

Third, more sophisticated radiative transfer models are needed to
analyse the spectra since the current column densities obtained from
slab models are only a crude representation of actual abundances and
their ratios. Retrieval models using a full 2D temperature and density
disk structure but a simplified chemistry is one possibility
\cite{Mandell12,Bosman17}. Retrieval using full chemical models is
computationally prohibitive \cite{Woitke19}, but such models can then
be used as inspiration for abundance distributions and search for
trends with physical parameters.  In the longer run, full
thermo-chemical disk models developed for individual sources can also
serve as testbeds for these more simplified approaches.  The
distribution of small versus large dust and presence of substructures
will be a key element in such models.

Finally, the consequences of the observed chemical diversity for the
composition of any planets forming in these disks need to be
investigated. Ultimately, these disk surveys need to be complemented
by JWST surveys of the chemical composition of the atmospheres of
mature planets to address the question what sets their composition.

\section*{Author Contributions}
Investigation: Grant, Tabone, van Gelder, Francis, Beuther, Bettoni,
Gasman, Arabhavi, Nazari, Vlasblom; Writing-original draft:
van Dishoeck; Writing-review: all authors; Software: Tabone, Grant,
van Gelder, Francis, Arabhavi, Bettoni, Gasman, Klaassen, Christiaens;
Visualization: Grant, Nazari, Francis, Vlasblom; Supervision: van
Dishoeck, Beuther, Kamp, Henning; Funding: van Dishoeck, Kamp,
Henning.

\section*{Conflicts of interest}
There are no conflicts to declare.

\section*{Acknowledgements}
The authors would like to thank the MIRI GTO JOYS and MINDS teams, as
well as the entire MIRI European and US instrument team.  Support from
StScI is also appreciated. The following National and International
Funding Agencies funded and supported the MIRI development: NASA; ESA;
Belgian Science Policy Office (BELSPO); Centre Nationale d’Etudes
Spatiales (CNES); Danish National Space Centre; Deutsches Zentrum fur
Luftund Raumfahrt (DLR); Enterprise Ireland; Ministerio De
Economi{\'a} y Competividad; Netherlands Research School for Astronomy
(NOVA); Netherlands Organisation for Scientific Research (NWO);
Science and Technology Facilities Council; Swiss Space Office; Swedish
National Space Agency; and UK Space Agency.

EvD acknowledges support from ERC Advanced grant 101019751 MOLDISK,
the Danish National Research Foundation through the Center of
Excellence “InterCat” (DNRF150), and DFG-grant 325594231, FOR 2634/2.
BT is a Laureate of the Paris Region fellowship program, which is
supported by the Ile-de-France Region and has received funding under
Marie Sklodowska-Curie grant agreement No. 945298.  DG 
thanks the Research Foundation Flanders for co-financing the present
research (grant number V435622N). TH is grateful for support from
the ERC Advanced grant 832428 Origins; IK, MvG, LF, AA and EvD from
TOP-1 grant 614.001.751 from the Dutch Research Council (NWO); and IK
from H2020-MSCA-ITN- 2019, grant no. 860470 (CHAMELEON).



\balance


\bibliography{biblio_evd} 
\bibliographystyle{rsc} 

\end{document}